\documentclass[11pt,onecolumn,amssymb,nofootinbib]{revtex4}
\usepackage{amsmath, amsthm, amscd, amssymb}
\usepackage{graphicx, braket}
\usepackage{bm}
\usepackage{bbm}
\usepackage{epstopdf}

\begin{document}

\title{\bf  Formulas for light deflection by a central mass,  solar system relativity tests, and modification of the lens equation, for a Weyl scaling invariant dark energy} \bigskip

\author{Stephen L. Adler}
\email{adler@ias.edu} \affiliation{Institute for Advanced Study,
Einstein Drive, Princeton, NJ 08540, USA.}

\begin{abstract}
We estimate the changes in light deflection and other solar system tests of general relativity when ``dark energy'' arises from a Weyl scaling invariant action, using the modified Schwarzschild-like spherically symmetric solution obtained in this case by Adler and Ramazano\u glu.  We show that the standard results of general relativity are modified by amounts that are far below current experimental errors.  However, because the metric is not conformally flat when the central mass $M$ vanishes, there is a contribution to light deflection proportional to the cosmological constant $\Lambda$, with opposite sign to the  one proportional to  $M$. This could have implications for gravitational lensing in cosmological contexts, so we have  calculated  the corresponding correction to the ``lens equation'' for gravitational lensing.   We also give formulas for the light deflection angle through order $\Lambda M$, assuming a  parametrized spherically symmetric metric for the external region of the central mass.

\end{abstract}

\vfill\eject

\maketitle
\section{Introdution}

\subsection{Weyl scaling invariant dark energy action}

In a series of papers over the last nine years,   as reviewed in \cite{A1} (with a detailed discussion of motivations), we have studied consequences of the postulate that ``dark energy'' arises from an action constructed from
nonderivative metric components so as to be invariant under the Weyl scaling $g_{\mu\nu}(x) \to \lambda(x) g_{\mu\nu}(x)$, where $\lambda(x)$ is a general scalar function.  This novel dark energy action is given by
\begin{equation}\label{dark}
S_{\rm dark~energy}=-\frac{\Lambda}{8 \pi G} \int d^4x  ({}^{(4)}g)^{1/2}(g_{00})^{-2}~~~,
\end{equation}
where $\Lambda$ is the cosmological constant, $G$ is Newton's constant, and ${}^{(4)}g=-\det({g_{\mu\nu})}$.  The action of Eq. \eqref{dark} is compatible with a 3+1 foliation of spacetime associated with the cosmic microwave background rest frame, and is three-space, although not four-space, generally covariant.  Equation \eqref{dark}  reduces to the usually assumed dark energy action,
\begin{equation}\label{darkold}
S_{\rm dark~energy}=-\frac{\Lambda}{8 \pi G} \int d^4x  ({}^{(4)}g)^{1/2}~~~,
\end{equation}
in homogeneous cosmological contexts with $g_{00}(x)\equiv 1$, so only gives rise to cosmological effects when there are inhomogeneities.

\subsection{The remainder of the gravitational action}

In applying  Eq.\eqref{dark} we assume that the part of the gravitational action that is second order and higher in metric derivatives is given by the usual Einstein-Hilbert action, which is four-space generally covariant.  Thus, we are not taking a requirement of full general covariance of the gravitational action is a guiding principle.  Instead, we note that it is now widely believed that both the Einstein-Hilbert action, and the ``cosmological constant'' action describing the observed acceleration of the expansion of the universe, are effective actions emerging at long wavelengths from an underlying theory that unifies gravitation and quantum field theory. This underlying theory may even be a discrete, rather than a continuum theory.  If this surmise is correct, there is no reason to mandate that the different segments of the gravitational effective action should obey the same symmetry principles; it is possible for the Einstein-Hilbert action to be four-space general covariant, while the action for the nonderivative components of the metric is Weyl scaling invariant, but only three-space general coordinate invariant.

This raises a number of questions that we briefly address here. (i)  First, we still assume that matter actions, and specifically the action for the electromagnetic field, are constructed to obey the equivalence principle, by replacing the flat spacetime Minkowski metric with the gravitational metric $g_{\mu\nu}$,   with no direct couplings to the assumed dark energy density.  Then massive and massless particles still follow geodesic trajectories, since as shown by Weinberg \cite{weinequiv},  this is an immediate consequence of the equivalence principle and makes no reference to the structure of the gravitational part of the action . (ii)  Second, as we have noted in our earlier papers, and was also discussed previously by Jacobson and Mattingly \cite{jacob}, introducing a frame dependence into the action results in the energy-momentum tensor derived by the usual four-space recipe being non-covariantly-conserved, which would lead to  inconsistent Einstein equations.  However, rather than seeing this as a fundamental obstacle to frame dependence as in \cite{jacob}, we have adopted the 3+1 foliation procedure of getting the spatial components of the energy-momentum tensor by varying the total action with respect to the spatial components $g_{ij}$ of the metric, and then getting the space-time and time-time components of the energy-momentum tensor by solving the equations for covariant conservation.  This procedure of ``covariant completion'', discussed in detail in \cite{A1}, has been successfully applied to stationary spherically symmetric and axially symmetric metrics by solving algebraic equations, and to the Friedmann-Lema\^itre-Robertson-Walker (FLRW) cosmological metric  by solving first order differential equations, and we believe that it is of general applicability.  The resulting covariantly conserved energy-momentum tensor can then be used as a source for the Einstein equations, with the  Bianchi identities fully respected.  (iii) Since the usual argument for absence of scalar gravitational waves uses full general coordinate invariance, scalar wave propagation has to be reexamined when the dark energy action is given by Eq. \eqref{dark}.  This requires a detailed calculation using the methods of perturbation theory around a FLRW background, undertaken in \cite{adlerscalar}, with the result that three-space general coordinate invariance of the action suffices to show that there are no propagating physical scalar gravitational waves, while tensor waves have the expected properties.\footnote{That is, before breaking temporal diffeomorphisms,  a putative propagating scalar mode can be removed by two gauge degrees of freedom, one temporal and one spatial.  After breaking temporal diffeomorphisms, the remaining spatial gauge mode still suffices to remove a propagating scalar degree of freedom, so such a mode is pure gauge and not physical.}

\subsection{Application to central masses}

Black holes as modified by the action of Eq.  \eqref{dark}
differ from their standard general relativity form within a distance of order $2 \times 10^{-18} (M/M_\odot)^2$cm from a nominal horizon radius $2M$, with $M$ the black hole mass and $M_\odot$ the solar mass.    Moreover,  detailed calculations in the spherically symmetric case show that as a consequence of the $(g_{00})^{-2}$ factor in Eq. \eqref{dark}, which would become infinite if $g_{00}$ were to vanish,  these modified, ``Schwarzschild-like'' black holes have neither an event horizon \cite{F}  nor an apparent horizon \cite{A2}, which leads to a number of interesting consequences as discussed in \cite{A1}.

Since general relativity is now very accurately tested, a natural question to ask is how the novel dark energy action of Eq. \eqref{dark} affects the usual solar system tests such as deflection of light near the sun, the perihelion shift of Mercury, and radar echo time delays.  In these applications of the Schwarzschild-like metric, $M$ will be the solar mass rather than a black hole mass.   One purpose of the new work in the present paper is to estimate the modifications to the usual formulas for solar system tests arising from Eq. \eqref{dark}, and to show that they are far below the level set by current error bounds.  But since our calculations show that the action of Eq. \eqref{dark} leads to a new order $\Lambda$ term in the standard formula for the light deflection angle, which could have implications for gravitational lensing at cosmological distances, we calculate the corresponding modification to the gravitational lens equation.  Additionally, we calculate the order $\Lambda M$ term for light deflection in a general parametrized spherically symmetric line element, which encompasses the line elements following from both Eq. \eqref{dark} and  Eq. \eqref{darkold}, and compare with prior results in the literature.

\section{The Schwarzschild-like metric to leading order in $\Lambda$}

In \cite{F} we computed, in a perturbation expansion in powers of $\Lambda$,  the Schwarzschild-like metric arising when the dark energy action of Eq. \eqref{dark} is combined with the usual Einstein-Hilbert gravitational action.  The result takes the following form in standard polar coordinates,  in geometrized units with the gravitational constant $G$ and the velocity of light $c$ set equal to unity,
\begin{align}\label{metric}
ds^2=& B(r)dt^2-A(r)dr^2 -r^2(d\theta^2 + \sin^2\theta d\phi^2)~~~,\cr
A(r)=&\exp\big(2H(r)\big)~~~,\cr
B(r)=&\exp\big(2F(r)\big)~~~,\cr
F(r)=&F^{(0)}(r)+\Lambda F^{(1)}(r)+O(\Lambda^2)~~~,\cr
H(r)=&H^{(0)}(r)+\Lambda H^{(1)}(r)+O(\Lambda^2)~~~,\cr
F^{(0)}(r)=&\frac{1}{2}\log(1-2M/r)~~~,\cr
H^{(0)}(r)=&-\frac{1}{2}\log(1-2M/r)~~~.\cr
\end{align}
Using Mathematica to integrate the equations for the first order perturbations $F^{(1)}(r)$ and $H^{(1)}(r)$, we obtained lengthy expressions given in Eqs. (34) and (37) of \cite{F}.

For $F^{(1)}$ we found, dropping terms here of order $M^2$ and higher that are not needed for what follows,
\begin{align}\label{ffinal}
F^{(1)}=& -\left\{\frac{1}{2}(r-2M)^2 +10 M (r-2M)+...  \right\}~~~,\cr
\end{align}
 and the corresponding expression for $H^{(1)}$ is
\begin{equation}\label{hfinal}
H^{(1)}=-\left\{(r-2M)^2-\frac{(1/2)r^4}{(r-2M)^2}+10M(r-2M)+...
\right\}~~~.
\end{equation}
Note that whereas $F^{(0)}(r)+H^{(0)}(r)=0$, the first order perturbations do not sum to zero, unlike the case for the Schwarzschild-de Sitter (SdS) metric (also called the Kottler metric)
\begin{equation}\label{SDS}
A(r)^{-1}=B(r)=1-2M/R -(1/3)\Lambda r^2~~~,
\end{equation}
that follows when the dark energy action is assumed to have the standard form of Eq. \eqref{darkold}, and unlike
generalizations of the SdS metric discussed by Zhang in \cite{Z}.

Since in  applications to solar system tests of general relativity $M$ will be the solar mass $M_\odot \simeq 1.5 \times 10^3 {\rm m}$, and $r$ values of interest will be much larger than this,  we can treat both  $\Lambda$ and $M$ as small quantities.  So we approximate $F^{(1)}$ and $H^{(1)}$ by their leading terms for $r>>M$,
\begin{align}\label{leadingfg}
F^{(1)}\simeq &-(C_B/2) r^2~~~,\cr
H^{(1)} \simeq &-(C_A/2) r^2~~~,\cr
C_A=&C_B=1~~~.\cr
\end{align}
The corresponding first order approximations to $A(r)$ and $B(r)$ are given by\footnote{With $C_A=C_B=1$, the $M=0$ expansions through order $\Lambda^2 r^4$ are  $A=1-\Lambda r^2 -\Lambda^2 r^4/5$ and $B=1-\Lambda r^2 +\Lambda^2 r^4/5$, so the formulas of Eq. \eqref{leadingab} should be useful for estimates when $\Lambda r^2$ is less than unity. We calculated the expansions to order $\Lambda^2$ using Eqs. (22)-(27) of \cite{F}.}
\begin{align}\label{leadingab}
A(r)\simeq &1+\frac{2M}{r}-C_A \Lambda r^2~~~,\cr
B(r)\simeq& 1-\frac{2M}{r}-C_B\Lambda  r^2~~~,\cr
\end{align}
and for general $C_A$ and $C_B$ as well as $C_A=C_B=1$,  $A(r)$ and $B(r)$ are not inverses of one another.   On the other hand, for the SdS metric one has $C_B=1/3~,~~C_A=-1/3$,~$C_A+C_B=0$, reflecting the fact that in this case  $A(r)^{-1}=B(r)$.

\section{Deflection of starlight to order $\Lambda$}

A formula for the deflection of starlight by the sun for general metric coefficient functions  $A(r)$ and $B(r)$ is conveniently given by Weinberg    \cite{wein}.  Letting $r_0$ be the distance of  closest approach of the light ray to the sun (which equals the solar radius for a grazing ray), and
$r_{\rm obs}$ and $r_{\rm star}$ the radii from the sun of the observer and the star being observed (which are taken as $\infty$ in \cite{wein}), the deflection angle $\Delta \phi$ is given by
\begin{equation}\label{defl}
\Delta \phi = \phi(r_0)- \phi(r_{\rm obs}) + \phi(r_0) - \phi(r_{\rm star}) -\pi~~~.
\end{equation}
The partial angular change $\phi(r_0)- \phi(R)$ for a light ray traversing from $r_0$ to $R$ is given by
\begin{equation}\label{phifn}
\phi(r_0)- \phi(R)=\int_{r_0}^R \frac{dr}{r} A^{1/2}(r)\left[\left(\frac{r}{r_0}\right)^2 \left(\frac{ B(r_0)}{B(r)}\right)-1\right]^{-1/2}~~~.
\end{equation}
Substituting Eqs. \eqref{leadingab} into Eq. \eqref{phifn} and simplifying through first order in small quantities, we get
\begin{equation}\label{phifn1}
\phi(r_0)- \phi(R)=\int_{r_0}^R \frac{dr}{r(r^2/r_0^2-1)^{1/2}} \left[1+\frac{M}{r}+\frac{Mr}{r_0(r+r_0)}-\frac{1}{2}(C_A+C_B) \Lambda r^2\right] ~~~.
\end{equation}
For the terms not containing $\Lambda r^2$, the upper limit $R$ of the integral can be extended  to $\infty$ when $R/r_0>>1$, giving the
the standard result for the deflection angle given in \cite{wein},
\begin{equation}\label{standard}
\Delta \phi_{\rm standard} = 4M/r_0~~~.
\end{equation}
The term containing $\Lambda r^2$ is easily integrated,
\begin{equation}\label{lambdaterm}
\int_{r_0}^R \frac{dr}{r(r^2/r_0^2-1)^{1/2}} (-r^2)=-r_0 (R^2-r_0^2)^{1/2}   ,
\end{equation}
giving an extra contribution to the deflection angle,
\begin{equation}\label{extra}
\Delta \phi_\Lambda= -(1/2) (C_A+C_B) \Lambda r_0 [(r_{\rm obs}^2-r_0^2)^{1/2}+(r_{\rm star}^2-r_0^2)^{1/2}]~~~,
\end{equation}
so that
\begin{equation}\label{total}
\Delta \phi_{\rm total}= \Delta \phi_{\rm standard}+\Delta \phi_\Lambda=4M/r_0~-(1/2) (C_A+C_B) \Lambda r_0 [(r_{\rm obs}^2-r_0^2)^{1/2}+(r_{\rm star}^2-r_0^2)^{1/2}].
\end{equation}
Thus when $C_A=C_B=1$, corresponding to the dark energy action of Eq. \eqref{dark},  we have
\begin{equation}\label{onecase}
\Delta \phi_{\rm total}= 4M/r_0- \Lambda r_0 [(r_{\rm obs}^2-r_0^2)^{1/2}+(r_{\rm star}^2-r_0^2)^{1/2}]~~~,
\end{equation}
showing that the $\Lambda$ term leads to a reduction in the deflection angle $\Delta \phi$.\footnote{
The reason there can be a nonzero deflection arising from spatially homogeneous dark energy when the central mass $M$ vanishes, is that Eq. \eqref{total} still refers to spherical coordinates around the origin $r=0$, which is a coordinate system that is not space translation invariant.}
On the other hand,  in the SdS case when $C_A+C_B=0$, the  modification to the usual answer is zero, reflecting the result  noted by Islam \cite{islam} that in this case there is no light deflection arising from the $\Lambda$ term by itself.  This is a consequence of the fact that when $M=0$ the SdS line element is conformally flat, and so propagation of null rays is unaffected.

Putting in numbers, in geometrized units the cosmological constant is $\Lambda \simeq 1.3 \times 10^{-52}\,{\rm m}^{-2}$, the solar radius is $r_0\simeq 7\times 10^{8}\,{\rm m}$, and taking the furthest distance to the observed  star as the galactic diameter so that $r_{\rm star} \sim 10^{21} \,{\rm m}>>r_{\rm obs}$, we get from Eq. \eqref{extra} the estimate
\begin{equation}\label{est1}
\Delta \phi_\Lambda \sim 10^{-22}~~~.
\end{equation}
This is far below the level of the standard answer $\Delta \phi_{\rm standard}\simeq 10^{-5}$ obtained by substituting the solar mass $M_\odot \simeq 1.5 \times10^3\, {\rm m}$ into    Eq. \eqref{standard}, which has been verified by observations to an accuracy of order a percent.  Finally, we note that for a star a galactic diameter away, $\Lambda r_{\rm star}^2 \sim 10^{-10}<<1$, justifying the use of a first order expansion in $\Lambda$ in calculating the dark energy modification.

\section{Precession of perihelia and radar echo delays}

For tests of general relativity involving solely the inner planets Earth and Mercury, with  distances from the sun of order $10^{11}\, {\rm m}$, we can use a simpler method to estimate an upper bound on the effect of $\Lambda$ on the standard results.  Writing Eq.
\eqref{leadingab} as
\begin{align}\label{leadingab1}
A(r)\simeq &1+\frac{2M}{r}[1-C_A\Lambda r^3/(2M)]~~~,\cr
B(r)\simeq& 1-\frac{2M}{r}[1+C_B\Lambda  r^3/(2M)]~~~,\cr
\end{align}
we see that the ``smallness parameter'' governing the relative size of the $\Lambda$ correction to the standard results is
\begin{equation}\label{smallness}
\Lambda r^3/(2M) ~~,
\end{equation}
which for $r \sim 10^{11} \,{\rm m}$ and $M=M_\odot \simeq 1.5 \times 10^3 {\rm m}$ is of order  $10^{-22}$.
This is far below the $O(10^{-4})$ accuracy to which the general relativistic extra perihelion  precession of Mercuary has been measured, and the $O(10^{-1})$ accuracy to which radar echo time delays have been measured.\footnote{We could not  use this method to place an upper bound on light deflection  since in that case,  for a star a galactic diameter away one has $\Lambda r^3/(2M_\odot )\sim 10^8>>1$, so instead we resorted to  an expansion of the light deflection formula from  \cite{wein} to first order in $\Lambda$.}

Similar  ``smallness parameter'' estimates can be made for systems at larger scales.  For a galaxy cluster, with $M \sim
10^{15} M_\odot$ and $r \sim  2 {\rm Mpc} \sim 6 \times 10^{22} {\rm m}$, we get $\Lambda r^3/(2M )\sim 10^{-2}<<1$.  But for a galaxy supercluster, with $M \sim  10^{16} M_\odot$ and $r \sim 2\times 10^{24} {\rm m}$, one has $\Lambda r^3/(2M) \sim 40$. So at the supercluster scale, one might see differences between the dark energy actions of
Eqs. \eqref{dark} and \eqref{darkold} having observable effects.

\section{Deflection angle to order $\Lambda M$}

Since the order $\Lambda$ correction to the light deflection angle given in Eq. \eqref{extra} grows with the impact parameter $r_0$, the observer radius $r_{\rm obs}$, and the star radius $r_{\rm star}$, it could be significant when these radii are large.  In lensing calculations $M$ can also be large, so we compute in this section the order $\Lambda M$ correction term to the deflection angle, as obtained from the formula given in \cite{wein}.   We continue to assume that $\Lambda r^2$ and $M/r$ are still small enough so that we can drop terms of order $\Lambda^2$ and $M^2$ and higher. We start from a spherically symmetric line element $ds^2=B(r) dt^2-A(r) dr^2 -r^2(d\theta^2 + \sin^2\theta d\phi^2)$, with $A(r)$ and $B(r)$ parametrized now
according to
\begin{align}\label{paramdef1}
A(r)=&1+2M/r-C_A \Lambda r^2 +D_A \Lambda M r+...,\cr
B(r)=&1-2M/r-C_B \Lambda r^2 +D_B \Lambda M r+...,\cr
\end{align}
with Table I giving the numerical values of the parameters $C_A,\,C_B,\,D_A,\,D_B$ for the line elements arising from the conventional and the Weyl scaling invariant choices of dark energy action.   Continuing to take $r_0$ as the radius of closest approach of the light ray to the
central mass, the deflection angle is still given by Eq. \eqref{defl}, and the partial angular change is still given by Eq. \eqref{phifn}, which we repeat here in the form
\begin{align}\label{phifnnew}
\phi(r_0)- \phi(R)=&\int_{r_0}^R \frac{dr}{r} Q(r)~~~,\cr
 Q(r)=&A^{1/2}(r)\left[\left(\frac{r}{r_0}\right)^2 \left(\frac{ B(r_0)}{B(r)}\right)-1\right]^{-1/2}~~~.\cr
\end{align}
Substituting the formulas of Eq. \eqref{paramdef1} into the formula for $Q(r)$ and expanding through order $\Lambda M$, we get after
some algebra (which we did ``by hand'' and then checked with Mathematica) the result
\begin{align}\label{final}
Q(r)=&(r^2/r_0^2-1)^{-1/2} [1 + M \left( \frac{1}{r}+\frac{r}{r_0(r_0+r)}\right) -\frac{1}{2}(C_A+C_B)\Lambda r^2 + \Lambda M X(r)]~~~,\cr
X(r)=&\frac{1}{2}(C_A + D_A -3C_B) r + \frac{1}{2} D_B \frac{r^2}{r_0+r} - \frac{1}{2} (C_A+C_B) \frac{r^3}{r_0 (r_0+r)}~~~.\cr
\end{align}
To perform the integrals, we first scale out $r_0$ by making the change of variable $r=r_0 u$, giving after rearranging the second and third terms of $X(r)$,
\begin{align}\label{phifnnew1}
\phi(r_0)- \phi(R)=&\int_{1}^{R/r_0} \frac{du}{u}(u^2-1)^{-1/2}
\left[1+\frac{M}{r_0}\left(\frac{1}{u}+\frac{u}{1+u}\right)
-\frac{1}{2}(C_A+C_B) \Lambda r_0^2 u^2 + \Lambda M r_0 X[u]\right]
~,\cr
 X[u]=&\frac{1}{2}(C_A + D_A -3C_B) u + \frac{1}{2} D_B u^2 - \frac{1}{2} (C_A+C_B+D_B) \frac{u^3}{1+u}~~~.\cr
\end{align}
The integrals over the terms of order 1 and order $M,\, \Lambda$ were already given in Sec. III.  The integral over the order $\Lambda M$ term gives
\begin{equation}\label{final1}
\int_{1}^{R/r_0} \frac{du}{u}(u^2-1)^{-1/2}
 \Lambda M r_0 X[u]= \Lambda M r_0 J[R/r_0]~~~,
 \end{equation}
 with $J[U]$ given by\footnote{In integrating to get Eq. \eqref{JJ}, the following identity is helpful:
${\rm ArcSinh}[((U-1)/2)^{1/2}]=(1/2){\rm ArcSinh}[(U^2-1)^{1/2}]$.  This is proved by using the formula ${\rm ArcSinh}(x)={\rm Log}[x+(x^2+1)^{1/2}]$ together with the identity $U+(U^2-1)^{1/2}=(1/2)[(U+1)^{1/2}+(U-1)^{1/2}]^2$.}
\begin{align}\label{JJ}
J[U]=&\frac{1}{2}(C_A + D_A -3C_B) J_1[U] + \frac{1}{2} D_B J_2[U] - \frac{1}{2} (C_A+C_B+D_B)J_3[U]~~~,\cr
J_1[U]=&\int_{1}^{U}du \,(u^2-1)^{-1/2}={\rm ArcSinh}[(U^2-1)^{1/2}]~~~,\cr
J_2[U]=&\int_{1}^{U}du \,(u^2-1)^{-1/2} u= (U^2-1)^{1/2}~~~,\cr
J_3[U]=&\int_{1}^{U}du \,(u^2-1)^{-1/2}\frac{u^2}{1+u}\cr
=&(U^2-1)^{1/2}+\left(\frac{U-1}{U+1}\right)^{1/2}-{\rm ArcSinh}[(U^2-1)^{1/2}]~~~.\cr
\end{align}
When $U=1$, the expression for $J[U]$ vanishes as required by Eq. \eqref{final1}.  From these equations we get the order $\Lambda M$ term in the deflection angle in terms of the parameters defined in Eq. \eqref{paramdef1}, with numerical values as given in Table I.

For the parameter values on the first line of Table I, $C_A=-C_B=-1/3, ~D_A=4/3,~D_B=0$, Eq. \eqref{final} gives $X(r)\equiv 0$, and so Eq. \eqref{JJ} evaluates to zero, which shows that in this case there is no $\Lambda M$  correction to the  deflection angle formula.    For the parameter values on the second line of Table I, corresponding to a Weyl scaling  invariant dark energy action, Eq. \eqref{JJ} evaluates to the nonzero value
\begin{equation}\label{JJ1}
  J[U]=-(U^2-1)^{1/2}+6\left(\frac{U-1}{U+1}\right)^{1/2}-12{\rm ArcSinh}[(U^2-1)^{1/2}]~~~.
\end{equation}

\begin{table} [ht]
\caption{Parameters $C_A,\,C_B,\,D_A,\,D_B$ for the spherically symmetric line element arising from the conventional and the Weyl scaling invariant dark energy actions.}
\centering
\begin{tabular}{c  c c c c c}
\hline\hline
dark energy type&Equations& $~~C_A~~$  & $~~C_B~~$  &$~~D_A~~$  &  $~~D_B~~$  \\
\hline
~~~~~conventional ~~~~~~~~~~& (2),(6)      &  -1/3  & 1/3  & 4/3    &  0  \\
Weyl scaling invariant & (1), (3)--(5)      &   1   & 1   &   -10   &  -14 \\
\hline\hline
\end{tabular}
\label{tab1}
\end{table}
\section{Discussion}

\begin{itemize}

\item  {\bf Vanishing of $\Lambda$ and $\Lambda M$ corrections in the SdS case.} As noted above, in the SdS case the $\Lambda$ contribution to light deflection vanishes, as observed by Islam
    \cite{islam}, and explained by the fact that when $M=0$ the SdS line element is de Sitter, which is conformally flat.   We also saw in the preceding section that in the SdS case the order $\Lambda M$ correction  vanishes as well.  To see that this agrees with results of other authors, we recall that $r_0$ in Weinberg's exposition \cite{wein} is not the ``impact parameter'', but rather the radius of closest approach of the light ray to the scattering center, determined by substituting $dr/dt=E=0$ into Eq. (8.4.19) of \cite{wein}, giving
    \begin{equation}\label{closest}
    J^2/r_0^2=1/B(r_0) = 1/(1-2M/r_0- \Lambda r_0^2/3)~~~,
    \end{equation}
    where $E$ and $J$ are the energy and angular momentum integration constants of the photon geodesic equation.  But the impact parameter $b$ in the absence of scattering is equal to the photon angular momentum $J$ (see Eq. (8.5.1) of \cite{wein} with photon velocity $V=1$, and Eq. (2) of Arakida and Kasai \cite{ara}.) So from Eq. \eqref{closest} we get
    \begin{equation}\label{br0relation}
    1/r_0=(1/b)(1-2M/r_0- \Lambda r_0^2/3)^{-1/2} \simeq (1/b)(1+M/r_0+  \Lambda r_0^2/6 )~~~,
    \end{equation}
    (which apart from a sign misprint in their $\Lambda$ term agrees with Eq. (14) of Bhadra, Biswas, and Sarkar \cite{bhadra}, and also agrees with Eq. (8) of Lebedov and Lake \cite{lebed}).
    Thus the deflection angle $4M/r_0$ that we calculated in terms of $r_0$, with no order $\Lambda M$ correction,
    becomes in terms of $b$,
    \begin{equation}\label{newdefl}
    4M/r_0=4M/b+ (2/3) \Lambda M  b + O(M^2) + O(\Lambda^2) ~~~.
    \end{equation}
    This is an equation given by many previous authors, including \cite{Z}, \cite{ara}, \cite{bhadra}, and \cite{lake}, as referenced in detail in \cite{Z}.  We see  that when $M/r_0<<1$ and $\Lambda r_0^2 <<1$, the difference between $r_0$ and $b$ is very small.

    \item  {\bf The first order correction when $C_A+C_B \neq 0$.}
    We have seen  that  modifications of the standard solar system tests of general relativity arising from the dark energy action of Eq. \eqref{dark} are much smaller than uncertainties in current measurements.  However, the calculation of light deflection leading to this conclusion showed that since $C_A+C_B =2$ for this action,  there is a nonzero correction term at first order in $\Lambda$, which vanishes for the standard  cosmological constant action where $C_A+C_B=0$. This correction term is given by Eq. \eqref{extra},
     which when $r_0$ is much smaller than $r_{\rm obs}$ and $r_{\rm star}$ is given by the simple formula
     \begin{equation}\label{extra1}
     \Delta \phi_\Lambda= - \Lambda r_0 (r_{\rm obs}+r_{\rm star})~~~,
     \end{equation}
     and acts to reduce the standard deflection angle  $4M/r_0$ calculated to leading order.
     The ratio of this term to the order $\Lambda M$ correction term calculated in Eq. \eqref{JJ1} is of order
     \begin{equation}\label{ratio1}
     \Lambda r_0 (r_{\rm obs}+r_{\rm star})/\big(\Lambda M r_0 (r_{\rm obs}+r_{\rm star})/r_0 \big)\simeq r_0/M~~~,
     \end{equation}
     which is typically very large, so the correction term of Eq. \eqref{extra1} is dominant.
     However, the ratio of Eq. \eqref{extra1} to the leading term $4M/r_0$ is
     \begin{equation}\label{ratio2}
      \Lambda r_0 (r_{\rm obs}+r_{\rm star})/(4M/r_0) = (1/4) \Lambda r_0^2 (r_{\rm obs}+r_{\rm star})/M~~~.
      \end{equation}
    Since gravitational lensing effects can involve very large star and observer radii from the lens, Eq. \eqref{ratio2} is not always very small. Thus it becomes relevant to compute the contribution of the order $\Lambda$ correction term to the lens equation.

    \item{\bf Correction to the lens equation}.  For a very clear discussion of lensing, which we shall follow here, see the article by Schneider in \cite{lenss}.  From this article we have adapted Fig. 1, which shows the typical geometry of a  gravitational lens system; we suggest that the reader peruse the following brief discussion in conjunction with pages 18-21 of \cite{lenss}.
         (For similar treatments of lensing, with the same basic notation, and similar figurs, see \cite{suyu}, \cite{dodel}, or \cite{lenss1}.)  In the notation of the just cited lensing literature, $r_{\rm obs}$ is termed $D_{\rm d}$, $r_{\rm star}$ is termed $D_{\rm ds}$, and $r_{\rm obs}+r_{\rm star}=D_{\rm d} + D_{\rm ds}$ is termed $D_{\rm s}$, and additionally $r_0$ is denoted by the magnitude $|\bm{\xi}|$.  The ``lens equation'' before including the $\Lambda$ correction reads
        \begin{equation}\label{lense1}
        \bm{\beta}=\bm{\theta}-\frac{D_{\rm ds}}{D_{\rm s}}\bm{\hat\alpha}( D_{\rm d}\bm{\bm{\theta }})
        \equiv \bm{\theta}-  \bm{\alpha} (\bm{\theta})~~~,
        \end{equation}
        where boldface greek letters denote 2 dimensional vector quantities oriented  perpendicular to the line of sight.
        With the addition of Eq. \eqref{extra1} to the deflection angle formula, the lens equation is modified to read
        \begin{equation}\label{lense2}
        \bm{\beta}=\bm{\theta}(1+\Lambda D_{\rm ds} D_{\rm d})-\frac{D_{\rm ds}}{D_{\rm s}}\bm{\hat \alpha}( D_{\rm d}\bm{\theta} )
        \equiv \bm{\theta}(1+\Lambda D_{\rm ds} D_{\rm d})   -  \bm{\alpha }(\bm{\theta})~~~,
        \end{equation}
        under the assumption that the surface mass density of the lens is invariant under spatial inversion.  This assumption should encompass a large class of astrophysical lenses; for asymmetric lenses that are not approximately inversion invariant, there will be additional shape-dependent contributions to Eq. \eqref{lense2}.

         Equation \eqref{lense2} is obtained by observing that the effect of the addition of Eq. \eqref{extra1} is to add $-\Lambda D_{\rm s} (\bm{\xi} -\bm{\xi^\prime})$ to the vectorized deflection angle with respect to an origin $\bm{\xi^{\prime}}$ in the planar projection of the lensing mass distribution.  If the lensing mass distribution is symmetric with respect to inversion around the center,  so that so that the surface mass density obeys  $\Sigma(\bm{\xi^\prime})=\Sigma(-\bm{\xi^\prime})$, the integral over the surface mass density of the $\bm{\xi^\prime}$ term in $-\Lambda D_{\rm s} (\bm{\xi} -\bm{\xi^\prime})$ vanishes, and the effect of Eq. \eqref{extra1} is to add $-\Lambda D_{\rm s} \bm{\xi}$ to the mass-averaged deflection angle  $\bm{\hat \alpha}(\bm{\xi})$ in Eq. (6) of \cite{lenss}.  After the rescaling $\bm{\xi}= D_{\rm d}\bm{\theta} $ of Eq. (7) of \cite{lenss}, which introduces angular coordinates, this gives an addition to the $\bm{\theta}$ term of the lens equation of
        \begin{equation}\label{finaladd}
        -\frac{D_{\rm ds}}{D_{\rm s}}(-\Lambda D_{\rm s} D_{\rm d}\bm{\theta}) =\Lambda D_{\rm ds} D_{\rm d}\bm{\theta}~~~,
        \end{equation}
        as shown in Eq. \eqref{lense2}.  In writing these equations we have assumed that $C_A+C_B=2$; for the case of general $C_A$, $C_B$, a factor of $(1/2)(C_A+C_B)$ should be inserted in front of $\Lambda$.

        To make a numerical estimate, suppose that $D_{\rm d} \sim D_{\rm ds} \sim 5 \times 10^9 ~{\rm light~years} \sim 5  \times 10^{25} {\rm m}$. Then $\Lambda D_{\rm ds} D_{\rm d} \sim 0.3$, a non-negligible correction.  Hence it will be of interest to study the consequences of the modified lens equation of Eq. \eqref{lense2} for cosmological applications of lensing.  This may give observational tests to distinguish between the dark energy actions of Eq. \eqref{dark}  and Eq. \eqref{darkold}.  When $\Lambda D_{\rm ds} D_{\rm d}$ approaches unity, there will be corrections to Eq. \eqref{finaladd} from order $\Lambda^2$ terms in the perturbation expansion, but still the order $\Lambda$ term computed here should give a useful qualitative and perhaps even semi-quantitative first estimate.\footnote{There is a long history in physics of leading order expressions proving useful beyond their formal range of applicability;  the $\epsilon$ expansion in renormalization group statistical mechanics, and the large-$N$ expansion in perturbative Yang-Mills theory, come to mind.}

        \item {\bf Corrections to the photon sphere radius and black hole shadow radius}. When the Event Horizon Telescope \cite{event} observes a black hole, what it sees is the black hole ``shadow'', which is the lensed image at infinity of the ``photon sphere''.   In collaboration with Virbhadra \cite{vir1} (where details are given),  we have calculated the photon sphere radius $r_{\rm ph}$ and the black hole shadow radius $r_{\rm sh}$,  using the parametrization of Eq. \eqref{paramdef1}, with the results
         \begin{align}\label{soln1}
           r_{\rm ph}\simeq & 3M [1 -(3/2) D_B \Lambda M^2]~~~,\cr
           r_{\rm sh}/r_{\rm ph}\simeq& 3^{1/2}[1- (3D_B-27C_B/2) \Lambda M^2]~~~.\cr
         \end{align}
            The current targets of the Event Horizon Telescope are the M87 black hole, with $M\sim 10^{13}\,{\rm m}$, and the Milky Way black hole SgA*, with $M\sim 10^{10}\,{\rm m}$. For these we find respectively $\Lambda M^2 \sim 10^{-26}$ and $\Lambda M^2 \sim 10^{-32}$, so the $\Lambda M^2$  corrections of Eq. \eqref{soln1}  are extremely small.

\end{itemize}

\vfill\eject

\section{Acknowledgement}
I wish to thank Zhen Zhang for correspondence in which he called my attention to his paper \cite{Z}, which stimulated the investigations reported here. I also wish to thank K. S. Virbhadra for correspondence and collaboration dealing with  cosmological constant corrections to the photon sphere and black hole shadow radii.  During final edits of this paper, I enjoyed the hospitality of the Aspen Center for Physics, which is supported by the National Science Foundation grant PHY-1607611.

Data availability statement:   Data sharing not applicable to this article as no datasets were generated or analysed during the current study.

Keywords:  cosmological constant; dark energy; Weyl scaling invariant; lens equation; light deflection; general relativity tests; Schwarzschild-like black hole

\vfill\eject
\begin{figure}[]
\begin{centering}
\includegraphics[scale=0.4]{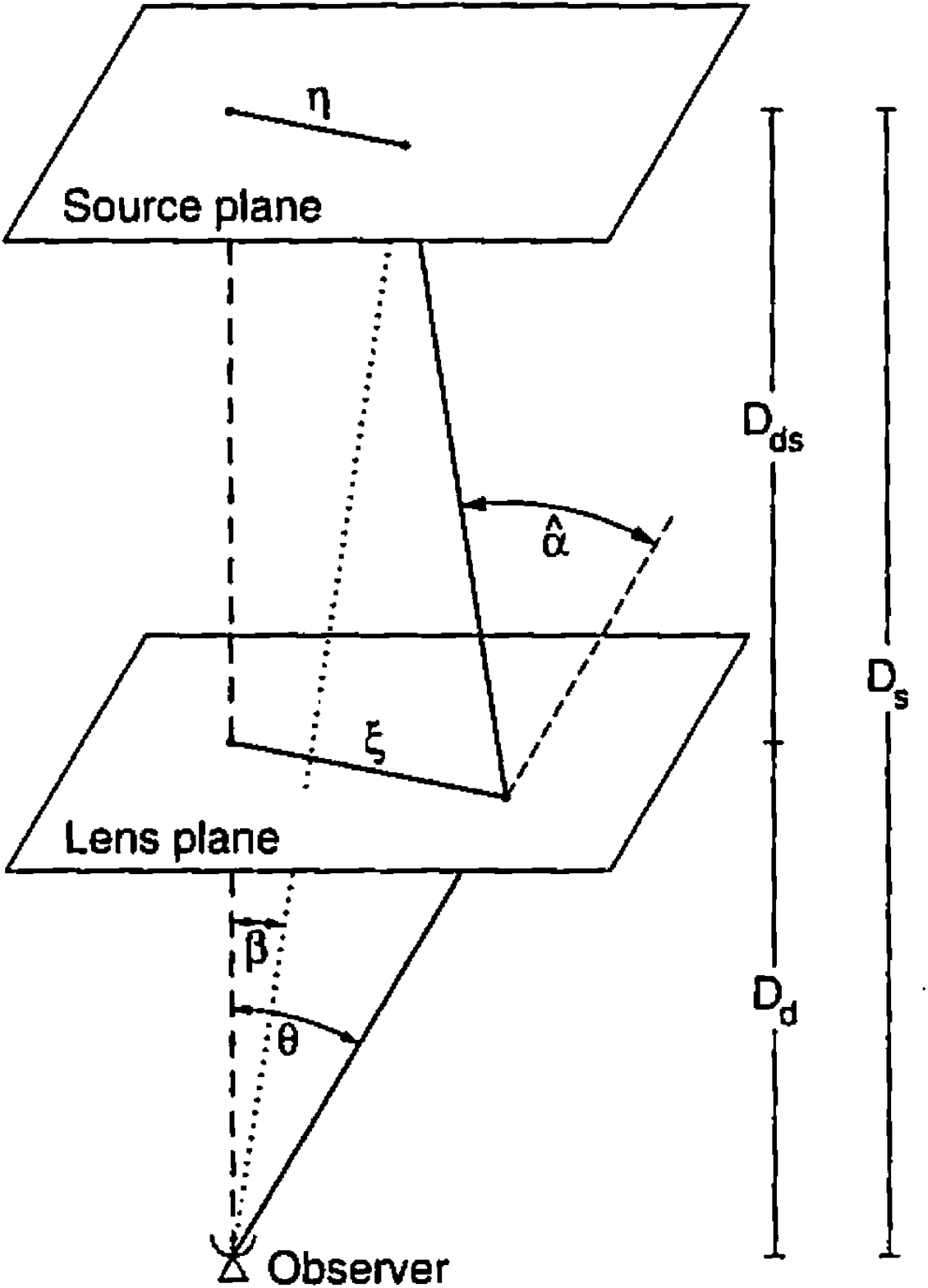}
\caption{ Geometry of a typical gravitational lens, from  \cite{lenss}, Fig. 12, p. 20 (reproduced with permission \copyright ESO).}
\end{centering}
\end{figure}

\end{document}